\documentclass[manuscript]{aastex}
\usepackage{graphicx,amsmath}

\begin{document}

\title{Cassini UVIS Observations of the Io Plasma Torus.
\\IV. Modeling Temporal and Azimuthal Variability}

\author{A.~J. Steffl} 

\affil{Southwest Research Institute, Department of Space Studies, 1050
  Walnut St., Suite 300, Boulder, CO, 80302}

\email{steffl@boulder.swri.edu}

\author{P.~A. Delamere, F. Bagenal}

\affil{University of Colorado, Laboratory for Atmospheric and Space
  Physics,Campus Box 392, Boulder, CO 80309-0392, USA.}


\begin{abstract}
  
  In this fourth paper in a series, we present a model of the
  remarkable temporal and azimuthal variability of the Io plasma torus
  observed during the {\it Cassini} encounter with Jupiter. Over a
  period of three months, the {\it Cassini} Ultraviolet Imaging
  Spectrograph (UVIS) observed a dramatic variation in the average
  torus composition. Superimposed on this long-term variation, is a
  10.07-hour periodicity caused by an azimuthal variation in plasma
  composition subcorotating relative to System III longitude. Quite
  surprisingly, the amplitude of the azimuthal variation appears to be
  modulated at the beat frequency between the System III period and
  the observed 10.07-hour period. Previously, we have successfully
  modeled the months-long compositional change by supposing a factor
  of three increase in the amount of material supplied to Io's
  extended neutral clouds. Here, we extend our torus chemistry model
  to include an azimuthal dimension. We postulate the existence of two
  azimuthal variations in the number of super-thermal electrons in the
  torus: a primary variation that subcorotates with a period of 10.07
  hours and a secondary variation that remains fixed in System III
  longitude. Using these two hot electron variations, our model can
  reproduce the observed temporal and azimuthal variations observed by
  {\it Cassini} UVIS.

\end{abstract}



\section{Introduction}

During the {\it Cassini} spacecraft's flyby of Jupiter (October 2000
through March 2001) the Ultraviolet Imaging Spectrograph (UVIS) made
extensive observations of the Io plasma torus. The sensitivity,
bandpass, resolution, and imaging capabilities of UVIS
\citep{Espositoetal04} coupled with the temporal coverage of the
observations make this a particularly rich dataset. Analysis and
modeling of these observations has led to remarkable new insights into
the behavior of the Io torus. In this paper we present the results of
our efforts to model the temporal and azimuthal variability of the Io
plasma torus observed by UVIS and discussed by \cite{Steffletal06a}.
To put this effort into proper context, here we recapitulate the prior
analysis and modeling of the {\it Cassini} UVIS observations.

Information about the UVIS Io torus dataset, including examples of the
observing geometry, images of the raw and processed data, and
descriptions of the data reduction and calibration techniques used was
presented in paper I \citep{Steffletal04a}. The analysis of a small
subset of UVIS Io torus observations obtained in January 2001, shortly
after the {\it Cassini} spacecraft's closest approach to Jupiter, was
presented in Paper II \citep{Steffletal04b}. Using the CHIANTI atomic
physics database \citep{Dereetal97, Youngetal03}, they derived the
composition and electron temperature of torus plasma as a function of
radial distance from Jupiter. The torus composition presented in Paper
II differs significantly from that derived from observations made in
the {\it Voyager} era \citep{Bagenal94}, with less oxygen and a lower
electron temperature. The CHIANTI-based spectral model of the Io torus
was used to analyze UVIS data obtained during a 45-day period on the
{\it Cassini} spacecraft's approach to Jupiter. Significant temporal
and azimuthal variations in torus composition were found, which were
presented in paper III \citep{Steffletal06a}.

To understand the processes governing the Io plasma torus and their
changes between the {\it Voyager} and {\it Cassini} epochs
\cite{Delamere:bagenal03} (hereafter referred to as DB03) developed a
physical chemistry model of the Io torus which builds on previous
modeling work by \cite{Barbosaetal83}, \cite{Shemansky88},
\cite{Barbosa94}, \cite{Schreieretal98}, and \cite{Lichtenbergetal01}.
The DB03 model, from which the azimuthal models discussed below are
directly descended, is a ``0-d'' or one-box model that calculates the
flow of mass and energy through one cubic centimeter of torus plasma
placed at a radial distance of 6~R$_J$. The effects of electron impact
ionization, recombination (both radiative and dielectronic), charge
exchange reactions, parameterized radial transport, Coulomb collisions
between species, and radiative energy losses on the torus plasma are
all included.

New mass is supplied to the torus in the form of neutral oxygen and
sulfur atoms which eventually become ions through either electron
impact ionization or charge exchange reactions. Conversely, mass is
lost from the model when an ion becomes neutralized through charge
exchange or recombination or through outward radial transport. The
rotation speed of plasma in the Io torus ($\sim$75~km/s at 6~$R_J$)
significantly exceeds Jupiter's escape velocity. When torus ions
becomes neutralized, the resulting atom is no longer constrained by
Jupiter's magnetic field and is quickly lost from the Io torus,
eventually forming an extended nebula hundreds of Jovian radii in size
\citep{Mendilloetal04}. Torus plasma is also convected radially
outwards by the interchange motions of magnetic flux tubes
\citep{Richardson:siscoe81, Siscoe:summers81}. However, the details
of convective radial transport are beyond the scope of the DB03 model;
instead, the lifetime of torus plasma against radial transport is
specified by an input parameter, $\tau_0$.

Energy is supplied to the torus via the ``pickup energy'' imparted to
new ions as they are accelerated from Keplerian orbital velocities to
near co-rotational with the magnetic field. As noted by
\cite{Shemansky88}, \cite{Smithetal88}, and others, pickup energy
alone cannot supply the torus with enough energy to maintain roughly
10$^{12}$~W of UV radiation, a thermal electron temperature of
$\sim$5~eV, and an average ionization state of $\sim$1.5, as are
observed: an additional source of energy is required. It is now widely
accepted that this additional energy is provided by a small population
of super-thermal electrons. This is supported by detections of a
high-energy component of the torus electron distribution function by
both the {\it Voyager} \citep{Scudderetal81,Sittler:strobel87} and
{\it Galileo} \citep{Frank:paterson00a} plasma instruments. In
addition, analysis of {\it Ulysses} URAP observations showed the
electron distribution function in the Io plasma torus resembles a
kappa distribution \citep{Meyer-Vernetetal95}.

In the torus model of DB03, the electron distribution function is
simplified as the sum of two Maxwellian populations: a thermal
population near the canonical 5~eV and a small (roughly 0.2\% of the
total electron density), hot population with temperature between
50--100~eV, consistent with \cite{Sittler:strobel87}. The temperature
of the hot population is held constant at a value specified by the
input parameter, $T_{e,hot}$. Since the hot electron population
rapidly couples to the thermal electron population via Coulomb
collisions (with a characteristic cooling time on the order of 30
minutes), maintaining a constant hot electron temperature requires the
hot population be continuously resupplied with energy. For the plasma
conditions observed by UVIS during the {\it Cassini} epoch, DB03 found
the hot electron source was responsible for up to 60\% (10$^{12}$ W)
of the total energy input to the torus.

The energy for these hot electrons must ultimately be derived from
Jupiter's rotation, but the particular details of the heating
mechanism remain poorly understood. \cite{Barbosa85} suggested that
the hot electrons may be heated by lower hybrid waves generated during
the thermalization of the pickup ion ring beam distribution. However,
if Io's extended neutral clouds are highly peaked near Io
\citep{Smyth:marconi03a, Burger:johnson04, Smyth:marconi05}, this
mechanism will produce a 13-hour periodicity (the synodic period
between Io and System III) which should be detectable in the {\it
  Cassini} UVIS observations. No such 13-hour periodicity was seen,
suggesting that ring beam thermalization is not the dominant
production mechanism.  \cite{Thorneetal97} reported rapid (100 km/s)
inward motion of hot tenuous plasma during a flux tube interchange
event. Such rapid motion could supply the Io torus with electrons that
have been heated in the outer torus/middle magnetosphere.
Alternatively, we propose that the hot electrons are produced locally
throughout the torus during small-scale flux tube interchange events.
These interchange events likely generate the field-aligned electron
beams detected by the {\it Galileo} PLS \citep{Frank:paterson00a} in
the shear region between inward and outward moving flux tubes.

An estimate of the total power available to the torus from these beams
of field-aligned electrons can be made using the following equation:

\begin{equation}
  P_{beam} = \Phi T f \pi(r_2^2-r_1^2)
\end{equation}

\noindent where $\Phi$ is the flux of electrons in the field-aligned 
beams, $T$ is the average electron energy, $f$ is the fraction of the
torus in which the beams occur, and $r_1$ and $r_2$ are radial
distance of the inner and outer edges of the torus.  From
\cite{Frank:paterson00a}, $\Phi \approx 10^8$~cm$^{-2}$~s$^{-1}$, $T
\approx 500$~eV (ranging between 100~eV and a few keV), and $f \approx
0.1$. Assuming a torus that extends from 6.0 to 7.5~R$_J$, yields a
total power input of a few 10$^{12}$~W, consistent with what is
required by DB03.

While a discussion of the origin of these beams is beyond the scope of
this paper, we note that a plausible heating mechanism is the
propagation of inertial Alfv\'en waves out of the plasma torus, as was
first discussed by \cite{Crary97} for the Io flux tube. We note that
the energy spectrum (100 eV to few keV) of the beams reported by
\cite{Frank:paterson00a} is consistent with electron acceleration to
roughly the Alfv\'en velocity just outside of the torus. For
additional discussion of acceleration by inertial Alfv\'en waves see
\cite{Suetal06} and \cite{Swift06}.

\subsection{Temporal variability during the {\it Cassini} era}

Analysis of the {\it Cassini} UVIS observations of the Io torus showed
significant changes in the composition of torus plasma between October
2000 and January 2001 \citep{Steffletal06a}. During this period, the
average mixing ratio (ion density divided by electron density) of
\ion{S}{2} in the torus declined by a factor of 2 during this period,
with a corresponding factor of 2 increase in the mixing ratio of
\ion{S}{4}. Observations by the {\it Galileo} Dust Detector System
(DDS) showed a dramatic increase, by over three orders of magnitude,
in the emission rate of Iogenic dust immediately prior to the UVIS
observations \citep{Kruegeretal03}. The enhanced dust emissions began
in July 2000, peaked in September 2000, and returned to pre-event
levels by December 2000. Presumably, this increase (the largest, by an
order of magnitude, observed by the {\it Galileo} DDS over a
seven-year period) was in response to some major volcanic event on Io,
possibly the eruption of Tvashtar Catena that deposited a Pele-like
ring of red material 1200~km in diameter on Io's surface between
February 2000 and December 2000 \citep{Geissleretal04}. A 400~km-high
plume over this region was observed by the {\it Cassini} Imaging
Science Subsystem in December 2000 \citep{Porcoetal03}.

These observations led \cite{Delamereetal04} to propose that an
increase in the amount of material supplied to the Io neutral clouds
might be responsible for the compositional changes seen by {\it
  Cassini} UVIS. Their model included a Gaussian increase in the
neutral source rate with an amplitude of 3.5 and a width of 22.5 days.
As the supply of neutrals increased, the densities in the model torus
rapidly increased to unrealistic levels, unless they were accompanied
by a corresponding decrease in the radial transport timescale, $\tau$.
Such an inverse relationship is expected if the radial transport of
plasma is driven by flux tube interchange \citep{Southwood:kivelson89,
  Brown:bouchez97, Pontiusetal98}. Since, the exact functional form of
the inverse relationship is dependent on the details of radial
convective transport, which remain poorly understood, a variety of
inverse relations were tested and a relation of $\tau \propto
\mathcal{S}_n^{-1}$ was adopted. In this work, we extend the model of
\cite{Delamereetal04} to include an azimuthal dimension.

\subsection{Azimuthal variability during the {\it Cassini} era}

Superimposed on the long-term temporal variation in torus composition,
UVIS observed a persistent azimuthal, i.e., longitudinal, asymmetry in
plasma composition, electron temperature, and equatorial electron
column density \citep{Steffletal06a}. This nearly-sinusoidal azimuthal
variation can be clearly seen in Fig.~\ref{mix_sinusoids_vs_time}.
The azimuthal variations of \ion{S}{2}, \ion{S}{3}, and electron
column density mixing ratios are all approximately in phase with each
other and are approximately 180$^{\circ}$ out of phase with the
variations of the mixing ratios of \ion{S}{4} and \ion{O}{2} and the
torus equatorial electron temperature.
\placefigure{mix_sinusoids_vs_time}

Over short timescales ($\lesssim$50 hours), the observed azimuthal
variability in the Io plasma torus is well described by a simple
sinusoidal curve with a period equal to the 9.925-hour System III
rotation period. However, the observed phase of the sinusoidal
variation slowly drifts to greater System III longitudes, at a rate of
12.5$^{\circ}$/day. This effect is clearly seen in the upper panel of
Fig.~\ref{uvis_phase_amp}. The rate of the phase increase implies that
the compositional variations observed by UVIS rotate Jupiter with a
period of 10.07 hours, 1.5\% longer than the System III rotation
period, yet 1.3\% shorter than the previously observed ``System IV''
period \cite{Sandel:dessler88, Brown95}. Careful analysis of the UVIS
observations using Lomb-Scargle periodograms \citep{Lomb76, Scargle82,
  Horne:baliunas86} confirmed strong torus periodicity with a period
of 10.07 hours (see Fig~\ref{datavsmodel_periodogram}) and a secondary
periodicity with a period close to the System III period of 9.925
hours.  \placefigure{uvis_phase_amp}

Numerous ground-based observations have established that plasma in the
Io torus lags rigid co-rotation with Jupiter's magnetic field
\citep{Brown83, Roesleretal84, Brown94b, Thomasetal01}. It is
therefore tempting to think of the 10.07-hour periodicity as direct
evidence of subcorotating plasma. However, spectroscopic measurements
of rotation speed of torus plasma have shown that the co-rotational
lag varies significantly as a function of radial distance from Jupiter
\citep{Brown94b, Thomasetal01}, whereas the phase drift seen in the
UVIS data remains coherent over a wide range of radial distances. A
similar argument was used by \cite{Brown95} to rule out plasma
subcorotation as the cause of the 10.21-hour ``System IV'' periodicity
seen in ground-based optical and {\it Voyager~1} UVS observations of
the Io plasma torus. Instead, the 10.07-hour periodicity in the UVIS
data, which is phenomenologically similar to the 10.21-hour ``System
IV'' periodicities, appears to be the result of a compositional wave
propagating azimuthally through the torus \citep{Brown94a, Steffl:phd,
  Steffletal06a}. Given the apparent similarities between the
10.07-hour periodicity in the UVIS data, the 10.224-hour periodicity
of \cite{Sandel:dessler88}, and the 10.214-hour periodicity of
\cite{Brown95}, we will subsequently refer to all three phenomena as
``System IV''.

In addition to the phase drift, the relative amplitudes of the
azimuthal variations in \ion{S}{2} and \ion{S}{4} mixing ratios vary,
in a roughly cyclical manner, between 5--25\%, as shown in the lower
panel Fig.~\ref{uvis_phase_amp}. The time between the observed peaks
of the \ion{S}{2} and \ion{S}{4} azimuthal variations is $\sim$29
days, suggestively close to the 28.8-day beat period between the
observed 10.07-hour {\it Cassini} epoch System IV period the
9.925-hour System III rotation period. Thus, the amplitude of the
azimuthal variation of these two ion species appears to be modulated
by the its location, relative to System III longitude. The amplitude
is greatest when the peak \ion{S}{2} mixing ratio is located near
$\lambda_{III}$=210$\pm$15$^{\circ}$ and smallest, i.e., most
azimuthally uniform, when the peak \ion{S}{2} mixing ratio is located
near $\lambda_{III}$=30$\pm$15$^{\circ}$. This effect is clearly seen
in \ion{S}{2} and \ion{S}{4}. However, the amplitude of the azimuthal
variation in the two primary ion species, \ion{O}{2} and \ion{S}{3},
was relatively constant during the UVIS observing period (in the range
of 2--5\%).

\section{Azimuthal Models \label{azimuthal_model_section}}

The one-box model of DB03 has five input parameters that can be
adjusted to match the conditions observed in the Io torus: the rate of
neutral atoms supplied to the torus ($\mathcal{S}_{n,0}$), the ratio
of oxygen to sulfur in the supplied neutrals ($O/S$), the fraction of
the total electron population heated to super-thermal levels
($f_{h,0}$), the temperature of the hot electrons ($T_{e,hot}$), and
the radial transport time scale ($\tau$). The sensitivity of the model
to these parameters is discussed in detail by DB03.

A successful azimuthal model must produce variations in torus
composition that i) are single-peaked and nearly sinusoidal, ii)
exhibit the temporal changes seen by {\it Cassini} UVIS over the
45-day approach phase, iii) drift, relative to System III longitude,
at a rate of approximately 12.5$^{\circ}$ day$^{-1}$, iv) vary in
amplitude with a period of 28.8 days, and v) have relative phases and
amplitudes consistent with those observed by UVIS. To meet these
requirements, we developed 1-d variants of the DB03 and
\cite{Delamereetal04} models, which are discussed below, in order of
increasing complexity.

\subsection{Basic Azimuthal Model}

In the basic azimuthal model, which serves as the basis for all
subsequent models, we extended the one-box model of DB03 to include 24
azimuthal bins, each corresponding to a 15$^\circ$ segment in System
III longitude, located at a radial distance of 6~R$_J$. The density
and temperature of torus plasma and neutral species are calculated for
each azimuthal bin, as are reaction rates between species. The model
plasma lags co-rotation with the System III coordinate system by an
amount, $\Delta v$, which can be a function of System III longitude.
Several models that incorporated an azimuthally variable co-rotational
lag were tested, but none could satisfactorily match the UVIS
observations. Therefore, we hold $\Delta v$ constant with location in
the torus. Spectroscopic observations of the Io torus have shown that
torus plasma at a radial distance of 6~R$_J$ lags co-rotation with the
magnetic field (i.e., the System III coordinate system) by 3--4~km/s
\citep{Brown94b, Thomasetal01}; we therefore adopt a value of $\Delta
v$=3.5~km/s, corresponding to a rotation period of 10.41 hours..

Io's extended neutral clouds are on Keplerian orbits of Jupiter. At
6$R_J$, they have a velocity of $\sim$57~km s$^{-1}$ relative to the
System III coordinate frame. Thus, neutrals will cross a 15$^\circ$
azimuthal bin in $\sim$2000~s. This is much faster than the
characteristic timescales for torus chemistry (ionization, charge
exchange, etc.), and it places an upper limit on the length of the
model time step. In the model, when a neutral atom becomes ionized, it
is given the plasma subcorotation velocity $\Delta v$ and pickup
energy of 380~eV for sulfur ions and 190 eV for oxygen ions.

The creation of Io's extended neutral clouds and their
three-dimensional spatial distribution are well beyond the scope of
this model \citep{Smyth:marconi03a, Burger:phd, Thomasetal04}.
Instead, the model starts with an azimuthally uniform neutral cloud.
New oxygen and sulfur atoms are added in an azimuthally uniform manner
at a rate controlled by the source parameters in
Eq.~\ref{delta_neutral_source_eqn}: $\mathcal{S}_{n,0}$, $\alpha_n$,
$t_n$, and $\sigma_n$. The rate at which neutral atoms are lost to the
torus, however, will generally vary as a function of azimuthal
position, resulting in neutral density variations of up to 20\%. We
assume the neutrals are confined to Jupiter's rotational equator with
a scale height of 0.5~R$_J$, consistent with \cite{Burger:phd}; our
models results are insensitive to changes in the neutral scale height.

The transport of mass and energy between azimuthal bins is handled via
a two-step Lax-Wendroff scheme \citep{Pressetal92} implemented after
the densities and temperatures for have been updated.  Following the
azimuthal transport, sinusoidal curves are fit to the azimuthal
variations of the model, in a procedure equivalent to that used in the
analysis of the UVIS data (cf. Steffl et al., 2004).

Torus plasma experiences a significant centrifugal force due to the
rapid rotation of Jupiter and finds an equilibrium about the position
on a given magnetic field line that is most distant from Jupiter's
rotation axis (see Bagenal, 1994). The locus of these points forms the
centrifugal equator, which is located 1/3 of the way between the
magnetic equator and the rotational equator \citep{Hilletal74,
  Cummingsetal80}. The offset between the centrifugal equator and
Jupiter's rotational equator varies as a function of System III
longitude, ranging from 0~R$_J$ at $\lambda_{III}$=20$^\circ$ and
$\lambda_{III}$=200$^\circ$ to 0.67~R$_J$ at
$\lambda_{III}$=110$^\circ$ and $\lambda_{III}$=290$^\circ$) at a
radial distance of 6~R$_J$.

Plasma pressure forces cause the torus plasma to spread out from the
centrifugal equator along magnetic field lines with a scale height
determined by the mass and temperature of the ions \citep{Bagenal94}.
Since typical scale heights of ions in the Io torus range between
1--2~R$_J$ \citep{Steffl:phd}, the number density of torus ions at the
rotational equator can vary by up to $\sim40\%$. Thus, timescales for
torus reactions are affected by the latitudinal density distribution,
with ion/neutral interactions affected more strongly than ion/ion
interactions.

Although our model includes only one spatial dimension (azimuthal), it
includes the effects of the latitudinal distribution of torus plasma
via the method of latitudinal averaging described by
\cite{Delamereetal05}. This method approximates the latitudinal
distribution of ions (neutrals) as a Gaussian centered on the
centrifugal (rotational) equator. The temperature distribution of each
species is assumed to be Maxwellian, so that temperature remains
constant with latitude. Reaction rates for each azimuthal bin are
determined by calculating the average reaction rate weighted by the
latitudinal distribution of the two reactants.

The DB03 model includes only the five major ion species in the torus:
\ion{S}{2}, \ion{S}{3}, \ion{S}{4}, \ion{O}{2}, and \ion{O}{3}.
Several other minor species such as \ion{Cl}{2} and \ion{Cl}{3}
\citep{Kueppers:schneider00, Feldmanetal01}, \ion{C}{3}
\citep{Feldmanetal04}, and \ion{S}{5} \citep{Steffletal04b} have been
detected in the torus. Additionally, the presence of Na and K ions is
inferred, given the observations of neutral Na and K near Io (see
review by Thomas et al., 2004). Finally, the torus plasma will also
contain protons. Although early work estimated the total flux tube
content consisted of 10--15\% protons, e.g., \citep{Tokaretal82}, more
recent work limits this value at only a few percent
\citep{Craryetal96, Wangetal98a, Wangetal98b, Zarkaetal01}. Although
we do not include any reactions involving these minor species, they
are included in our model's calculation of charge neutrality by
assuming that they compose 10\% of the total charge, such that:

\begin{equation}
  0.9N_e=S^+ + 2S^{2+}+ 3S^{3+} + O^+ + 2O^{2+}
\end{equation}

\noindent A similar equation for quasi-neutrality was used by 
\cite{Steffletal06a} in the analysis of the {\it Cassini} UVIS
spectra. Model results are generally insensitive to the assumed charge
fraction of these minor ion species, over the range of 0--20\%.

To accommodate the variations to the basic azimuthal model described
below, the fraction of hot electrons was allowed to vary as a function
of both time and location according to the following equation:

\begin{eqnarray}
  \label{fh_eqn}
  f_h(t,\lambda_{III}) & = & f_{h,0} \left(1+\alpha_h e^{-(t-t_h)^2/\sigma_h^2}\right)\\
  \nonumber & \times & \left(1 + 
    \alpha_{h,\lambda_{IV}} \cos (\lambda_{III} - \phi_{h,\lambda_{IV}}-\omega t) + 
    \alpha_{h,\lambda_{III}} \cos (\lambda_{III}-\phi_{h,\lambda_{III}}) \right)
\end{eqnarray}

\noindent However, in the basic model, $\alpha_h$,
$\alpha_{h,\lambda_{IV}}$, and $\alpha_{h,\lambda_{III}}$ are set to
zero.

\subsection{Time-Variable Model}

To reproduce the temporal changes in torus composition observed during
the {\it Cassini} epoch, we follow the method of \cite{Delamereetal04}
and include a Gaussian increase in the neutral source rate:

\begin{equation}
  \label{delta_neutral_source_eqn}
  \mathcal{S}_n (t)= \mathcal{S}_{n,0} \left(1+\alpha_n e^{-(t-t_n)^2/\sigma_n^2} \right)
\end{equation}

\noindent where $\mathcal{S}_{n,0}$ is the baseline value of the 
neutral source rate. Like \cite{Delamereetal04}, the date of the peak
neutral source, $t_n$, was held fixed at day 249 to match the center
of the increase in Iogenic dust flux observed by the {\it Galileo}
Dust Detector System \citep{Kruegeretal03}.

Latitudinal averaging of reaction rates produces an increase in the
characteristic timescales for torus reactions. As a result, although
the general trend of decreasing \ion{S}{2} and increasing \ion{S}{4}
seen in the UVIS data (c.f.  Fig.~\ref{datavsmodel_comp}) could be
reproduced, the rapidity of this change could not. Based on their
parameter sensitivity studies, DB03 noted that the hot electron
fraction ($f_h$) is the only model parameter capable of modifying the
torus composition on such rapid timescales (days). Therefore, a broad
Gaussian perturbation to the hot electron fraction was added to the
model by allowing $\alpha_h$ to be non-zero. $t_h$, and $\sigma_h$
were also allowed to vary to match the observed compositional change.

Adding a Gaussian increase in the hot electron fraction to match the
UVIS-observed composition is admittedly ad hoc. However, this increase
can be loosely justified by the following argument. As the the density
of the neutral clouds increases, there is increased mass loading in
the torus. The resulting increase in flux tube content, increases the
efficiency of outward radial transport of plasma, parameterized in our
model as $\tau$. This, in turn, will drive field-aligned currents that
could produce additional hot electrons.

\subsection{Subcorotating Hot Electron Model}
\label{subcorotating_elec_model}

The azimuthal variation in torus composition observed by {\it Cassini}
UVIS has a rotation period of 10.07 hours, slightly longer than the
9.925-hour System III rotation period. The torus plasma, however, has
an even longer rotation period (roughly 10.41 hours at 6~R$_J$), and
is a strong function of radial distance \citep{Brown94b,
  Thomasetal01}, so the subcorotation of the torus plasma can not be
directly responsible for the period of the azimuthal variations.
Instead, noting that the timescale for hot electrons to couple
energetically with the thermal electron population is of order tens of
minutes compared to timescales of several days or more for all other
torus processes (cf. Table~\ref{jan14_lifetime_table}), we introduce a
subcorotating variation in the fraction of hot electrons by allowing
$\alpha_{h,\lambda_{IV}}$ and $\phi_{h,\lambda_{IV}}$ in
Eq.~\ref{fh_eqn} to be non-zero. The angular velocity of the hot
electron variation relative to System III, $\omega$, is set to
12.5$^{\circ}$/day, corresponding to a rotation period of 10.07 hours.
We subsequently refer to this model as the ``Subcorotating Hot
Electron Model''.

\subsection{Dual Hot Electron Model}
\label{modulated_elec_model}

In addition to having a period of 10.07 hours, the amplitude of the
azimuthal variation in torus composition appears to change with time
in a roughly periodic way. The period of the amplitude variation (28.8
days) is identical to the beat period produced by the interference of
the observed 10.07-hour period and the System III period of 9.925
hours. A similar variation in the UV brightness of the torus was
observed by the {\it Voyager~2} UVS \citep{Sandel:dessler88}; the
14.1-day period of this brightness variation matches the beat period
between the {\it Voyager}-era System IV period of 10.22 hours and the
System III period. Further analysis by \cite{Yangetal91} concluded
that the System IV periodicity in the {\it Voyager} data was indeed
independent of System III and that the observed 14.1-day period was
likely the result of the interference of phenomena with the System IV
and System III periods.

To test whether a similar beat frequency modulation of torus
composition could be produced by super-thermal electrons, we added a
second hot electron variation that remains fixed in System III
coordinates by allowing $\alpha_{h,\lambda_{III}}$ to be non-zero.
This model, with all sixteen input parameters non-zero, will be
referred to as the ``Dual Hot Electron Model''.

\section{Model Results}

\subsection{Basic Azimuthal Model}

The five parameters of the basic azimuthal model ($\mathcal{S}_{n,0}$,
$O/S$, $f_{h,0}$, $T_{e,hot}$, and $\tau$) were adjusted to match the
azimuthally-averaged torus composition derived from the {\it Cassini}
UVIS observations of 2001 January 14 \citep{Steffletal04b}. The values
of these parameters are shown in Table~\ref{model_parameters_table}.
The final equilibrium state produced using these parameters is used as
the initial condition of the Io torus for the subsequent three models.
\placetable{model_parameters_table}

Characteristic timescales for torus loss processes are given in
Table~\ref{jan14_lifetime_table}, which illustrates the importance of
charge exchange reactions (and resonant charge exchange reactions in
particular) in the Io plasma torus. For example, while the primary
loss mechanism of neutral sulfur from the extended clouds is electron
impact ionization by the thermal ($\sim$5~eV) electron population,
neutral oxygen is lost primarily through resonant charge exchange with
\ion{O}{2}.  Likewise, the dominant loss process of \ion{S}{2} is not
ionization (by either the thermal or hot electron population), but
rather the resonant charge exchange reaction $S^+ + S^{++} \rightarrow
S^{++} + S^+$.  Although resonant charge exchange reactions do not
change the number densities of torus ions, they do redistribute energy
between ion species.  \placetable{jan14_lifetime_table}

The offset between the centrifugal and rotational equators produces a
slight azimuthal variation in torus composition, as can be seen in
Fig.~\ref{cm3_noaz}. Where the two equator planes intersect (at
110$^\circ$ and 290$^\circ$), \ion{S}{2} shows a 1\% increase,
relative to the average mixing ratio, due primarily to the increased
rate electron impact ionization of \ion{S}{1}. Approximately
30$^\circ$ downstream (torus plasma moves in the direction of
increasing System III longitude), \ion{S}{4} exhibits a 1\% decrease
due primarily to the increased rate of the charge exchange reaction
$S^{3+} + O \rightarrow S^{2+} + O^+$. However, this azimuthal
variation is double-peaked and clearly much smaller in amplitude than
the azimuthal variations observed by {\it Cassini} UVIS.
\placefigure{cm3_noaz}

\subsection{Time-Variable Azimuthal Model}

After including Gaussian perturbations to both the neutral source rate
and the hot electron fraction, the time-variable azimuthal model
reproduces the observed temporal behavior of the sulfur ion mixing
ratios, as seen in Fig.~\ref{datavsmodel_comp}.  The parameter values
used to produce this figure are listed in
Table~\ref{model_parameters_table}. However, the model fails to
reproduce the temporal behavior of \ion{O}{2}. This discrepancy could
arise from an error (or errors) in the rate coefficients for reactions
involving \ion{O}{2}. Alternatively, the ratio of oxygen to sulfur
atoms supplied to the extended neutral clouds may not be constant.
Regarding this latter possibility, \cite{Spenceretal00} report the
discovery of gaseous S$_2$ in Io's Pele plume at the level of
SO$_2$/S$_2$ = 3--12. If the putative volcanic event responsible for
the increase in the torus neutral source (the Tvashtar eruption of
2000) was sufficiently rich in S$_2$ (or SO), the ratio of oxygen to
sulfur atoms supplied to the neutral clouds could have temporarily
decreased. As the neutral source rate returned to pre-event levels,
the O/S ratio of the neutrals would rise, producing a gradual increase
in the mixing ratio of \ion{O}{2}.  \placefigure{datavsmodel_comp}

\subsection{Subcorotating Hot Electron Model}

As expected, the subcorotating sinusoidal variation in the hot
electron fraction produced single-peaked azimuthal variations in torus
composition with a rotation period of 10.07 hours. The top panel of
Fig~\ref{datavsmodel_pa_nosys3} shows the model azimuthal variation
drifts in System III longitude by 12.5$^\circ$ per day and roughly
matches the observed phase increase. The best-fit parameter values of
the subcorotating hot electron model are listed in
Table~\ref{model_parameters_table}.

Including an azimuthal variation in the hot electron fraction did not
change the average composition of the torus. This behavior, due to the
symmetric nature of the azimuthal perturbation (a sine wave), greatly
simplifies the fitting procedure. Since the temporal variation model
parameters ($\alpha_n$, $t_n$, $\sigma_n$, $\alpha_h$, $t_h$, and
$\sigma_h$) are decoupled from the azimuthal variation model
parameters ($\alpha_{h,\lambda_{IV}}$ and $\phi_{h,\lambda_{IV}}$),
once the best-fit temporal parameters have been determined only two
additional parameters need to be varied to match the phase increase of
the azimuthal variation in composition.

The interaction of the azimuthal variation in hot electron fraction
with the neutral source increase results in a dramatic increase (from
6\% to 22\%) in the relative amplitude of the \ion{S}{4} azimuthal
variation, centered around DOY 249. This increase is largely caused by
the efficient removal of \ion{S}{4} ions via the charge exchange
reaction $S^{3+} + O \rightarrow S^{2+} + O^+$ at longitudes where the
centrifugal and rotational equators intersect. During the {\it
  Cassini} UVIS approach observations (DOY 275--320), the amplitude of
the model \ion{S}{4} azimuthal variation monotonically decreases to
its pre-event level. Conversely, interaction with the increased
neutral source results in a slight decrease (from 12\% to 10\%) in the
amplitude of the model \ion{S}{2} azimuthal variation. During the {\it
  Cassini} UVIS observation period, the \ion{S}{2} variation exhibits
a slight maximum near day 295. The amplitude behavior of both the
\ion{S}{4} and \ion{S}{2} variations produced by the subcorotating hot
electron model is in stark contrast to the UVIS observations, as seen
in the bottom panel of Fig.~\ref{datavsmodel_pa_nosys3}.
\placefigure{datavsmodel_pa_nosys3}

\subsection{Dual Hot Electron Model}

Like the subcorotating hot electon model, the dual hot electron model
can produce azimuthal variations in torus composition that drift to
higher System III longitudes, but only if $\alpha_{h,\lambda_{IV}}
\gtrsim \alpha_{h,\lambda_{III}}$. If this condition is not satisfied,
the azimuthal variation in composition becomes fixed in System III
coordinates. In contrast to the subcorotating hot electron model, the
rate at which the azimuthal variation drifts is not constant. This
effect can be seen in the top panel of Fig.~\ref{datavsmodel_pa}. For
most of the time, the model azimuthal variation lags co-rotation with
a period of 10.02 hours, corresponding to an angular velocity in the
System III frame of $\omega = 8.3^\circ$/day. However, when the two
variations are roughly 180$^\circ$ out of phase, the rotation period
of the compositional variation increases to 10.23 hours ($\omega =
26.1^\circ$/day). The UVIS data show a similar trend, though the
difference in rotation period is less dramatic, changing from 10.02
hours ($\omega = 8.8^\circ$/day) to 10.11 hours ($\omega =
16.2^\circ$/day).  The changing rotation period of the compositional
variation is in no way related to the speed of the torus plasma, but
rather is a wave phenomenon produced by the interaction of the two hot
electron variations with the torus plasma.
\placefigure{datavsmodel_pa}

Although the instantaneous rotation period of the azimuthal variation
varies over the 28.8-day beat period, Lomb-Scargle periodogram
analysis shows strong periodicity at 10.07 hours, with secondary
periodicity at the System III period of 9.925 hours, just like the
UVIS data (see Fig.~\ref{datavsmodel_periodogram}). For a stable
neutral source ($\alpha_n = \alpha_h = 0$) and the {\it Cassini} epoch
nominal torus composition, the amplitude of the \ion{S}{2} variation
ranges from 5\% to 17\% over the 28.8-day beat period. For \ion{S}{4},
these values are 3\% and 10\%.  \placefigure{datavsmodel_periodogram}

As with the subcorotating hot electron model, the interaction of the
neutral source increase with the two azimuthal variations in hot
electron fraction results in a dramatic increase in the amplitude of
the variation of \ion{S}{4} (7\% variation at minimum amplitude to
32\% variation at maximum amplitude), while producing only a minimal
effect on the amplitude of the \ion{S}{2} variation. As the neutral
source rate gradually returns to its pre-event level, the amplitude of
the \ion{S}{4} variation also decreases, reaching its pre-event
amplitudes after approximately 70 days. Similar behavior can be seen
in the amplitude of the \ion{S}{4} variation seen by UVIS (see bottom
panel of Fig.~\ref{datavsmodel_pa}).

Figure~\ref{datavsmodel_lonprofile} shows the azimuthal variation in
torus composition produced by the dual hot electron variation model
during epochs of maximum (Day 279) and minimum (Day 293) amplitude
compared to the azimuthal variation observed by {\it Cassini} UVIS.
The model parameters used to produce the results shown in
Figs.~\ref{datavsmodel_periodogram}--\ref{datavsmodel_lonprofile} are
given in Table~\ref{model_parameters_table}.
\placefigure{datavsmodel_lonprofile}

\section{Discussion \label{Discussion_section}}

Given the simplifying assumptions, the match between the output of the
dual hot electron variation model and the {\it Cassini} UVIS
observations of the Io torus is remarkable. Like the observed
variations, the model variations are single-peaked and have an average
rotation period of 10.07 hours. The model variations show the same
relative phase and amplitude as the observed variations, exhibit a
28.8-day beat period, and, with the exception of \ion{O}{2}, match the
azimuthally-averaged torus composition as a function of time.

The large number of model input parameters (16), makes it difficult to
assess the uniqueness of our model solution, as a complete search
through the 16-dimensional parameter space is computationally
prohibitive. However, despite considerable effort, we found no other
region in the parameter space of our model that produced results that
match the UVIS observations. Regarding the uncertainty in parameter
values, models run with deviations of up to 10\% from the parameter
values given in Table~\ref{model_parameters_table} generally yield
results that are similar to those shown in Figs. \ref{datavsmodel_pa}
and \ref{datavsmodel_lonprofile}, whereas models with parameter
deviations greater than $\sim$10\% produce a notably poorer match to
the UVIS data.

The value of the model parameter $\phi_{IV}$ provides no real insight
into the nature of the torus; rather, this parameter merely represents
the phase of the subcorotating hot electron variation at an arbitrary
time, $t=0$ (in this case, chosen to be 2000-01-01). In contrast, the
requirement that $\phi_{III} = 290^\circ$ for the model to match the
UVIS data is significant. Since Jupiter's dipole magnetic field is
tilted toward a System III longitude of 200$^\circ$, the rotational
and centrifugal equator planes intersect at the System III longitudes
of 110$^\circ$ and 290$^\circ$. In the absence of any temporal
variations, this results in a 20\% increase in the pickup energy
supplied to the Io torus at these longitudes. By itself, this has very
little effect on the torus composition, as shown by
Fig~\ref{cm3_noaz}. If, as we propose, the super-thermal electrons in
the Io torus are primarily produced via field-aligned currents, the
increase in mass loading at these longitudes should drive additional
field-aligned currents resulting in an increase of hot electrons at
these longitudes. However, increased mass loading (and thus an
increased hot electron fraction) should occur at both
$\lambda_{III}=290^\circ$ and $\lambda_{III}=110^\circ$. Models using
a double-peaked System III-fixed hot electron variation were not able
to reproduce the amplitude variations seen in the UVIS data and
resembled the results of the Subcorotating Hot Electron model shown in
Fig.~\ref{datavsmodel_pa_nosys3}. Our best model requires a hot
electron maximum at $\lambda_{III}=290^\circ$, as expected, but a hot
electron minimum at $\lambda_{III}=110^\circ$. What causes the
symmetry between $\lambda_{III}=110^\circ$ and
$\lambda_{III}=290^\circ$ to be broken? One possibility is that
higher-order components of Jupiter's magnetic affect the conductivity
of the ionosphere in such a way that the field-aligned currents that
accelerate electrons in the torus are strongly favored at
$\lambda_{III}=290^\circ$ over $\lambda_{III}=110^\circ$.

While the System III-fixed variation in hot electrons likely has its
origin in the interaction between Jupiter's magnetic field and the
Jovian ionosphere, the source of the System IV periodicity in the Io
torus has been enigmatic. Previously, \cite{Dessler85} and
\cite{Sandel:dessler88} have proposed a secondary, high-latitude
component of Jupiter's magnetic field that lags co-rotation by a few
percent could be responsible for producing the System IV periodicity.
While neither the UVIS observations nor our efforts to model them can
rule out the existence of such a high-latitude component, this
hypothesis is problematic. Presumably, such a magnetic field component
would affect the Jovian aurora as well as the Io plasma torus.
However, the morphology of the Jovian aurora is strongly fixed in
System III longitude, and although the temporal coverage of the Jovian
aurora has been somewhat sporadic, there have been no auroral
phenomena reported at either the 10.2 or 10.07-hour periods typified
by System IV \citep{Clarkeetal04}. Furthermore, it is difficult to
reconcile a secondary magnetic field component with the changing
period of System IV phenomena: 10.224 hours during the {\it Voyager}
epoch \citep{Sandel:dessler88}, 10.214 hours in 1992 \cite{Brown95}
and 10.07 hours during the {\it Cassini} epoch \citep{Steffletal06a}.
We therefore consider the existence of a subcorotating high-latitude
magnetic field component improbable, though we are unable to offer a
satisfactory alternative.

As discussed above, the rotation speed of torus plasma is a strong
function of radial distance from Jupiter. Both the Subcorotating Hot
Electron Model and the Dual Electron Model use a corotational lag of
3.5~km/s. However, both models are insensitive to changes in the
amount of corotational lag, over the range observed in the torus
(0--4~km/s). In particular, the Dual Hot Electron Model can reproduce
the temporal and azimuthal variations in composition observed by UVIS,
regardless of the radial distance at which it is run, consistent with
the radial uniformity of the System IV period observed by
\cite{Brown95}.

\section{Conclusions}
During the {\it Cassini} spacecraft's flyby of Jupiter, the UVIS
instrument observed remarkable temporal and azimuthal variations in
the composition of the Io plasma torus.  The azimuthal variations,
which are primarily seen in the ion species \ion{S}{2} and \ion{S}{4},
lag co-rotation with the magnetic field and are decoupled from the
rotation speeds of both the torus plasma and Iogenic neutral clouds.
The strength of the azimuthal variation changes in a seemingly
periodic manner with a period of 28.8 days---the beat period between
System III and the observed 10.07-hour rotation period of the
azimuthal variations.

To model the temporal and azimuthal changes observed by UVIS, we have
extended the torus chemistry model of \cite{Delamere:bagenal03} to
include an azimuthal dimension. Our preferred model, which includes
two independent azimuthal variations in the amount of hot electrons in
the Io torus, one subcorotating and one fixed in System III, can
reproduce the UVIS observations remarkably well. The major findings of
this paper are summarized below.

\begin{enumerate}
  
\item The months-long change in the average composition of the Io
  plasma torus can be modeled by introducing a factor of 3 increase to
  the rate of oxygen and sulfur atoms supplied to the extended neutral
  clouds that are the source of the torus plasma coupled with a 30\%
  increase in the fraction of hot electrons in the Io torus. This
  result is similar to that reported by \cite{Delamereetal04}.
  
\item An azimuthal variation in the fraction of hot electrons in the
  Io torus that rotates with a period of 10.07 hours can produce
  subcorotating azimuthal variations in torus composition like those
  observed by {\it Cassini} UVIS.
  
\item The interference of the subcorotating hot electron variation
  with a second hot electron variation that remains fixed in System
  III can produce the beat frequency modulation in the amplitude of
  the azimuthal variations also seen by {\it Cassini} UVIS

\end{enumerate}

\appendix

\section{Model Reaction Rate Coefficients}

We describe here the sources of the various rate coefficients used in
our model.

\subsection{Ionization}
Rate coefficients for electron impact ionization are calculated using
the fit formulae given by \cite{Voronov97}. These fits are based on
the University of Belfast group recommended data
\citep{Belletal83,Lennonetal88}.

Since the lifetimes of neutral and ionic oxygen and sulfur against
photoionization are several orders of magnitude longer than the
characteristic timescales for other processes such as electron impact
ionization, charge exchange, and recombination \citep{Huebneretal92},
the effects of photoionization can be ignored.

\subsection{Recombination}
For ions with multiple electrons, recombination rate coefficients are
usually divided into two separate processes: radiative recombination
and dielectronic recombination \citep{Osterbrock89}.  The total
recombination rate coefficient is the sum of the radiative and
dielectronic recombination terms. Total recombination rate
coefficients for oxygen ion species are obtained from \cite{Nahar99}.
Total recombination rates for S$^{3+}$ $\rightarrow$ S$^{2+}$ and
S$^{2+}$ $\rightarrow$ S$^+$ are obtained from \cite{Nahar95} and the
associated erratum \cite{Nahar96}. In general, the rate coefficients
published by Nahar agree well with previously published results at low
temperatures. At temperatures typical of the Io torus, however, the
Nahar rates can be up to an order of magnitude lower than previously
published values. Since the work by Nahar is the most recent treatment
of the recombination rate problem for sulfur and oxygen ions and
employs a more sophisticated technique than previous studies, these
rates are used in the model. For the recombination of S$^+$
$\rightarrow$ S, the radiative recombination rate coefficient of
\cite{Shull:vansteenberg82a} is used, while the dielectronic
recombination rate coefficient is obtained from \cite{Mazzottaetal98}.

The recombination rate coefficients used in this work differ from
those used in previous versions of the torus chemistry model
\citep{Delamere:bagenal03, Delamereetal04, Delamereetal05}. However,
since recombination reactions are generally much slower than other
processes that occur in the torus (cf.
Table~\ref{jan14_lifetime_table}) these changes do not significantly
affect our conclusions.

\subsection{Charge Exchange}
Charge exchange reactions play an important role in the torus
chemistry. Seventeen charge exchange reactions between atomic and
ionic species of sulfur and oxygen, listed in Table~1 of DB03, are
included in the model. All charge exchange reaction rates are taken
from \cite{McGrath:johnson89}.

\subsection{Radiation}
The radiative rate coefficients of the model are obtained from the
CHIANTI atomic physics database version 4.2 \citep{Dereetal97,
  Youngetal03}.


\acknowledgements
\noindent {\bfseries Acknowledgments}\\
Analysis of the {\it Cassini} UVIS data was supported under contract
JPL~961196.


\clearpage

\begin{deluxetable}{llcccc}
  \tablewidth{0pt} 
  \tabletypesize{\scriptsize}
  \tablecaption{Best-fit Azimuthal Model Parameters \label{model_parameters_table}}
  \tablehead{\colhead{Parameter Description} & \colhead{Symbol} & \colhead{Basic} & 
    \colhead{Time Variable} & \colhead{Subcorotating Hot} & \colhead{Dual Hot} \\
    \colhead{} & \colhead{} & \colhead{Model} & 
    \colhead{Model} & \colhead{Electron Model} & \colhead{Electron Model}}
  \tablecolumns{6}
  \startdata
  Neutral source rate at the rotational equator (cm$^{-3}$ s$^{-1}$) &
    $\mathcal{S}_{n,0}$ & 9.9x10$^{-4}$ & 9.9x10$^{-4}$ & 9.9x10$^{-4}$ & 9.9x10$^{-4}$ \\  
  O to S ratio of neutral source & O/S & 1.55 & 1.55 & 1.55 & 1.55 \\
  Fraction of hot electrons & $f_{h,0}$ & 0.00235 & 0.00235 & 0.00235 & 0.00235 \\
  Temperature of hot electrons (eV) & $T_{e,hot}$ & 55 & 55 & 55 & 55 \\
  Radial transport timescale (days) & $\tau_0$ & 62 & 62 & 62 & 62 \\
  Amplitude of neutral source increase & $\alpha_n$ & 0.0 & 2.4 & 2.4 & 2.4 \\
  Date of neutral source increase (DOY 2000) & $t_n$ & \nodata & 249 & 249 & 249 \\
  Gaussian width of neutral source increase (days) & $\sigma_n$ & \nodata & 30 & 30 & 30 \\
  Amplitude of hot $e^-$ fraction increase & $\alpha_h$ & 0.00 & 0.30 & 0.30 & 0.30 \\
  Date of hot $e^-$ fraction increase (DOY 2000) & $t_h$ & \nodata & 279 & 279 & 279 \\
  Gaussian width of hot $e^-$ fraction increase (days) & $\sigma_h$ & \nodata & 60 & 60 & 60 \\
  Amplitude of System IV hot $e^-$ variation & $\alpha_{h,\lambda_{IV}}$  & 0.0 & 0.0 & 0.43 & 
    0.43\\
  Phase of System IV hot $e^-$ variation ($^\circ$) & $ \phi_{h,\lambda_{IV}}$ & \nodata & 
    \nodata & 60 & 60 \\
  Angular velocity between Systems III and IV ($^\circ$/day) & $\omega$ & \nodata & 
    \nodata & 12.5 & 12.5 \\
  Amplitude of System III hot $e^-$ variation & $\alpha_{h,\lambda_{III}}$ & 0.0 & 0.0 & 0.0 & 
    0.30 \\
  Phase of System III hot $e^-$ variation ($^\circ$) & $\phi_{h,\lambda_{III}}$ & \nodata & 
    \nodata & \nodata & 290 \\

  \enddata
\end{deluxetable}

\begin{deluxetable}{lccccccc}
  \tablewidth{0pt} 
  \tabletypesize{\scriptsize}
  \tablecaption{Characteristic Timescale of Torus Loss Processes \label{jan14_lifetime_table}}
  \tablehead{\colhead{Loss Mechanism} & \colhead{S I} & \colhead{S II} & \colhead{S III} & 
    \colhead{S IV} & \colhead{O I} & \colhead{O II} & \colhead{O III}}
  \startdata
   Thermal e$^-$ impact ionization & 0.8 & 16.0 & 463 & 10400 & 6.4 & 926 & 70700 \\
   Hot e$^-$ impact ionization &  15.9 & 43.0 & 128 & 338 & 43.5 & 168 & 438 \\
   Recombination & \nodata & 1410 & 324 & 123 & \nodata & 4050 & 1330 \\
   S$^{+}$ + S$^{++}$ $\rightarrow$ S$^{++}$ + S$^{+}$& \nodata & 3.0 & 10.4 & \nodata & \nodata & \nodata & \nodata \\
   S + S$^+$ $\rightarrow$ S$^+$ + S$^\ast$ & 5.0 & 85.2 & \nodata & \nodata & \nodata & \nodata & \nodata \\
   S + S$^{++}$ $\rightarrow$ S$^+$ + S$^+$ & 105 & \nodata & 6240 & \nodata & \nodata & \nodata & \nodata \\
   S + S$^{++}$ $\rightarrow$ S$^{++}$ + S$^\ast$ & 4.0 & \nodata & 240 & \nodata & \nodata & \nodata & \nodata \\
   S + S$^{+++}$ $\rightarrow$ S$^{+}$ + S$^{++}$ & 14.0 & \nodata & \nodata & 142 & \nodata & \nodata & \nodata \\
   O + O$^{+}$ $\rightarrow$ O$^{+}$ + O$^\ast$ & \nodata & \nodata & \nodata & \nodata & 2.6 & 43.3&    \nodata \\
   O + O$^{++}$ $\rightarrow$ O$^{+}$ + O$^{+}$ & \nodata & \nodata & \nodata & \nodata & 627 & \nodata & 1070 \\
   O + O$^{++}$ $\rightarrow$ O$^{++}$ + O$^\ast$ & \nodata & \nodata & \nodata & \nodata & 60.4& \nodata & 104 \\
   O + S$^{+}$ $\rightarrow$ O$^{+}$ + S$^\ast$ & \nodata & 8510 & \nodata & \nodata & 1990 & \nodata & \nodata \\
   S + O$^{+}$ $\rightarrow$ S$^{+}$ + O$^\ast$ & 10.8 & \nodata & \nodata & \nodata & \nodata & 734 & \nodata \\
   S + O$^{++}$ $\rightarrow$ S$^{+}$ + O$^{+}$ & 13.9 & \nodata & \nodata & \nodata & \nodata & \nodata & 95.7 \\
   S + O$^{++}$ $\rightarrow$ S$^{++}$ + O$^{+}$ + $e^-$ & 20.1 & \nodata & \nodata & \nodata & \nodata & \nodata & 138 \\
   O + S$^{++}$ $\rightarrow$ O$^{+}$ + S$^{+}$ & \nodata & \nodata & 205 & \nodata & 13.8& \nodata & \nodata \\
   O$^{++}$ + S$^{+}$ $\rightarrow$ O$^{+}$ + S$^{++}$ & \nodata & 262 & \nodata & \nodata & \nodata & \nodata & 105 \\
   O + S$^{+++}$ $\rightarrow$ O$^{+}$ + S$^{++}$ & \nodata & \nodata & \nodata & 24.4 & 9.6 & \nodata & \nodata \\
   O$^{++}$ + S$^{++}$ $\rightarrow$ O$^{+}$ + S$^{+++}$ & \nodata & \nodata & 376 & \nodata & \nodata & \nodata & 43.4 \\
   S$^{+++}$ + S$^{+}$ $\rightarrow$ S$^{++}$ + S$^{++}$ & \nodata & 585 & \nodata & 346 & \nodata & \nodata & \nodata \\
   Radial transport & \nodata & 62.0 & 62.0 & 62.0 & \nodata & 62.0 & 62.0 \\
     \hline
   Total of all loss processes & 0.5 & 2.2 & 7.3 & 12.8 & 1.3 & 20.9 & 12.5 \\
  \enddata   
  \tablecomments{Characteristic timescales given are for the
    ``nominal'' {\it Cassini} epoch torus composition and have been
    azimuthally averaged. All timescales have units of days.}

\end{deluxetable}

\begin{figure}
  \includegraphics[scale=.75]{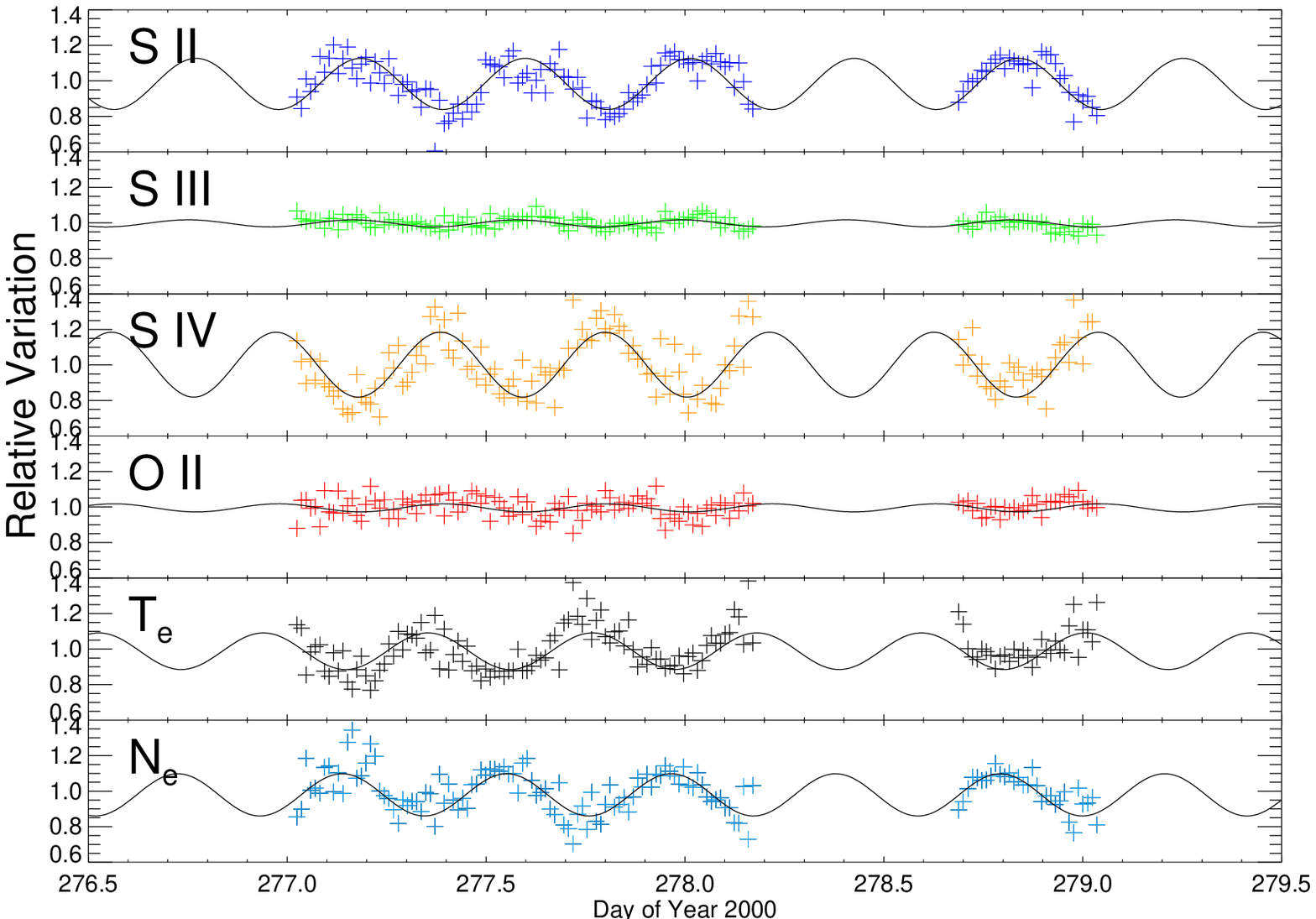}
  \caption[]{Steffl et al. \label{mix_sinusoids_vs_time} \\
    Relative ion mixing ratios, electron temperature, and electron
    column density for a typical 3-day period obtained from the dusk
    ansa. Values are normalized to the average value over the 3-day
    period. The best-fit sinusoids for this period are overplotted.
    Note the strong anti-correlation of \ion{S}{2} with \ion{S}{4} and
    equatorial electron temperature with equatorial electron column
    density. Figure taken from \cite{Steffletal06a}.}
\end{figure}

\clearpage
\noindent {\bfseries Fig.~\ref{mix_sinusoids_vs_time} Caption} \\
Relative ion mixing ratios, electron temperature, and electron column
density for a typical 3-day period obtained from the dusk ansa. Values
are normalized to the average value over the 3-day period. The
best-fit sinusoids for this period are overplotted.  Note the strong
anti-correlation of \ion{S}{2} with \ion{S}{4} and equatorial electron
temperature with equatorial electron column density. Figure taken from
\cite{Steffletal06a}.
\\

\begin{figure}
  \includegraphics[scale=.6]{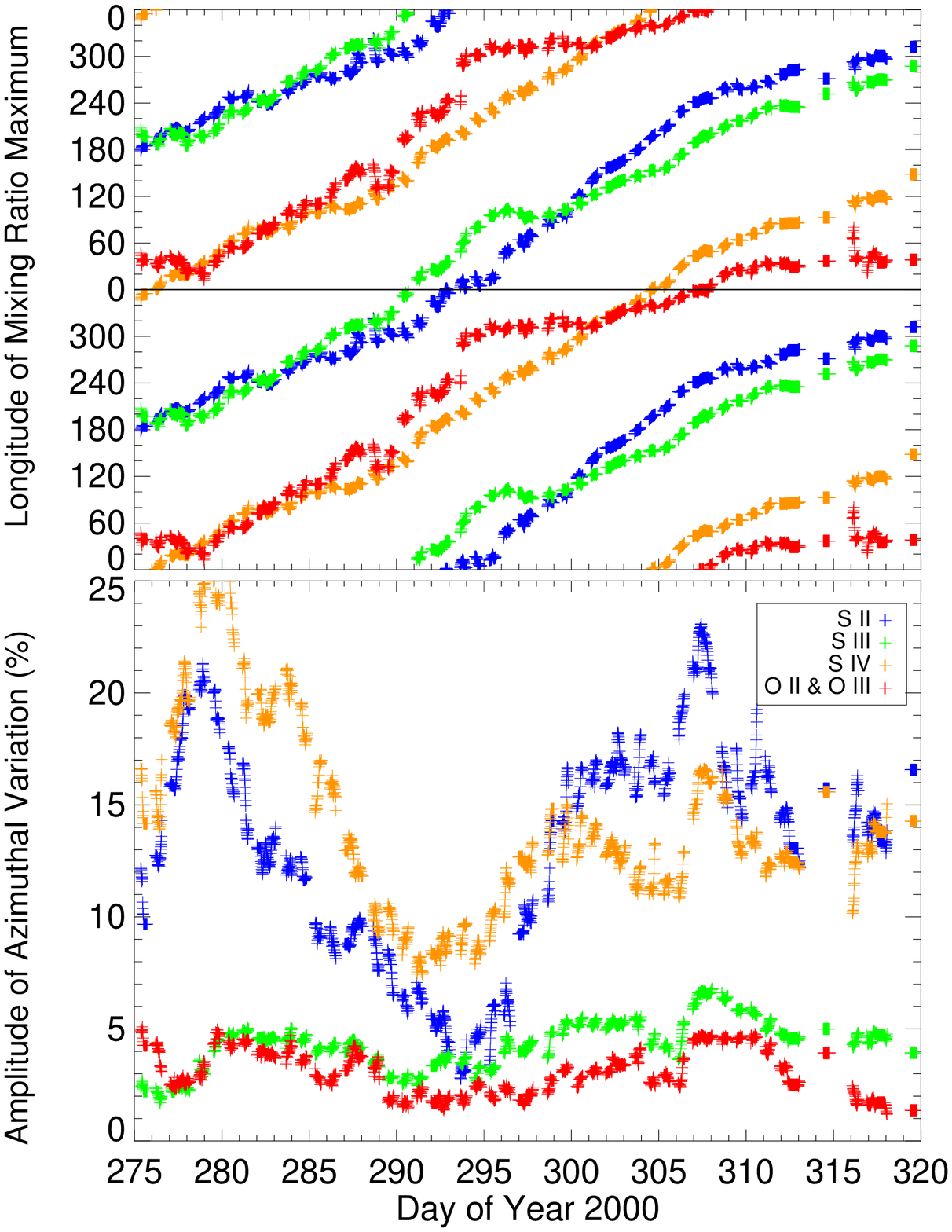}
  \caption[]{Steffl et al. \label{uvis_phase_amp} \\ 
    Azimuthal variations in the Io plasma torus as observed by {\it
      Cassini} UVIS. The top panel shows the location (in System III
    coordinates) of the peak in mixing ratio of the primary ion
    species in the Io torus as a function of time. All four ion
    species show a roughly linear trend of increasing phase with time.
    The bottom panel shows the relative amplitude (as a percentage) of
    the azimuthal variations as a function of time. The relative
    amplitudes of \ion{O}{2} and \ion{S}{3} remain around the few
    percent level, while the comparatively less abundant ion species
    \ion{S}{2} and \ion{S}{4} vary between 4--25\%. Adapted from
    \cite{Steffletal06a}.}
\end{figure}

\noindent {\bfseries Fig.~\ref{uvis_phase_amp} Caption} \\ 
Azimuthal variations in the Io plasma torus as observed by {\it
  Cassini} UVIS. The top panel shows the location (in System III
coordinates) of the peak in mixing ratio of the primary ion species in
the Io torus as a function of time. All four ion species show a
roughly linear trend of increasing phase with time.  The bottom panel
shows the relative amplitude (as a percentage) of the azimuthal
variations as a function of time. The relative amplitudes of
\ion{O}{2} and \ion{S}{3} remain around the few percent level, while
the comparatively less abundant ion species \ion{S}{2} and \ion{S}{4}
vary between 4--25\%. Figure taken from \cite{Steffletal06a}.
\\

\begin{figure}
  \includegraphics[scale=.75]{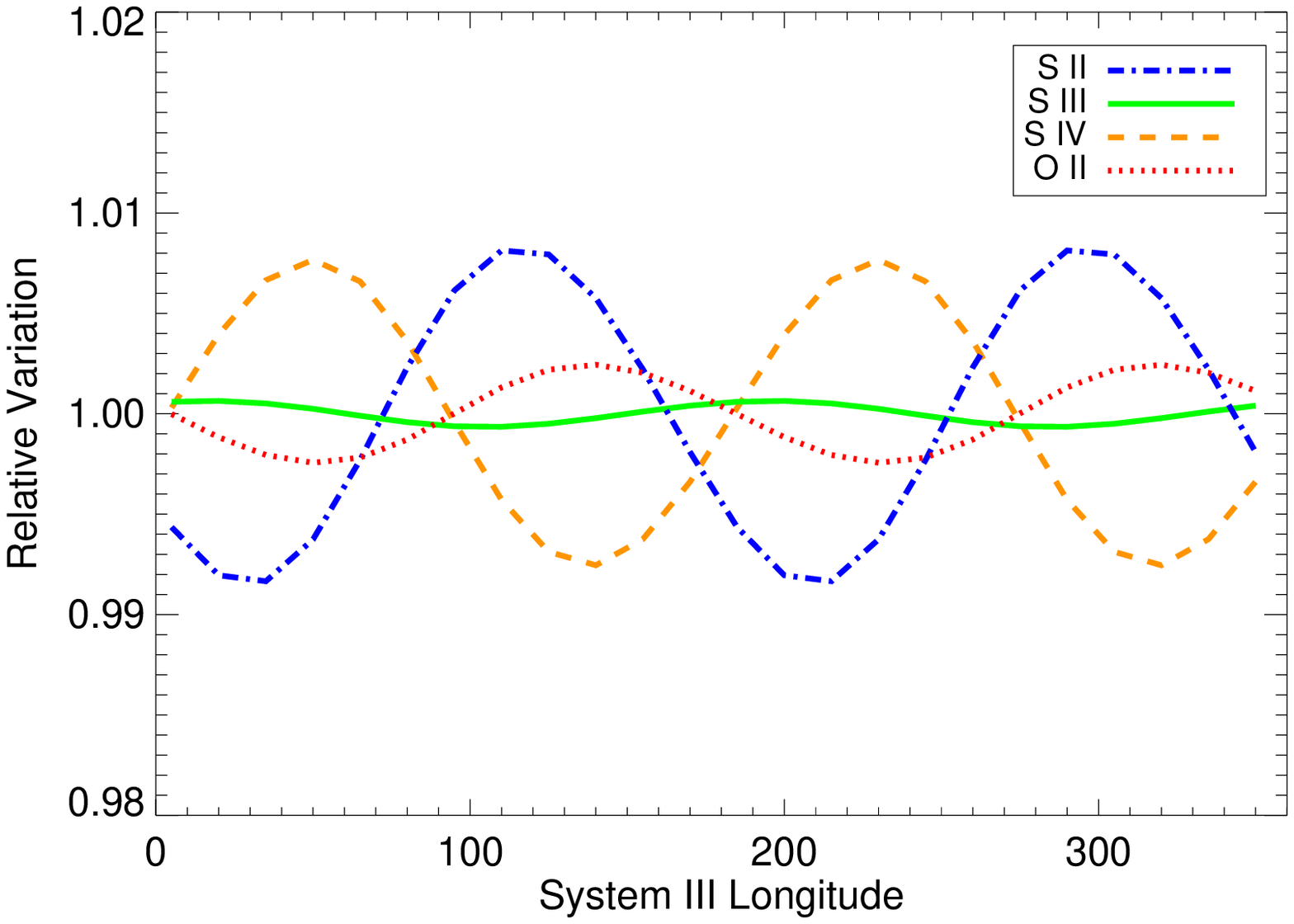}
  \caption[]{Steffl et al.  \label{cm3_noaz} \\
    Relative variation in torus composition produced by the basic
    azimuthal model. The small azimuthal variations are caused by the
    offset between the centrifugal and rotational equators and are
    therefore double-peaked.}
\end{figure}

\noindent {\bfseries Fig.~\ref{cm3_noaz} Caption} \\
Relative variation in torus composition produced by the basic
azimuthal model. The small azimuthal variations are caused by the
offset between the centrifugal and rotational equators and are
therefore double-peaked.
\\

\begin{figure}
  \includegraphics[scale=.75]{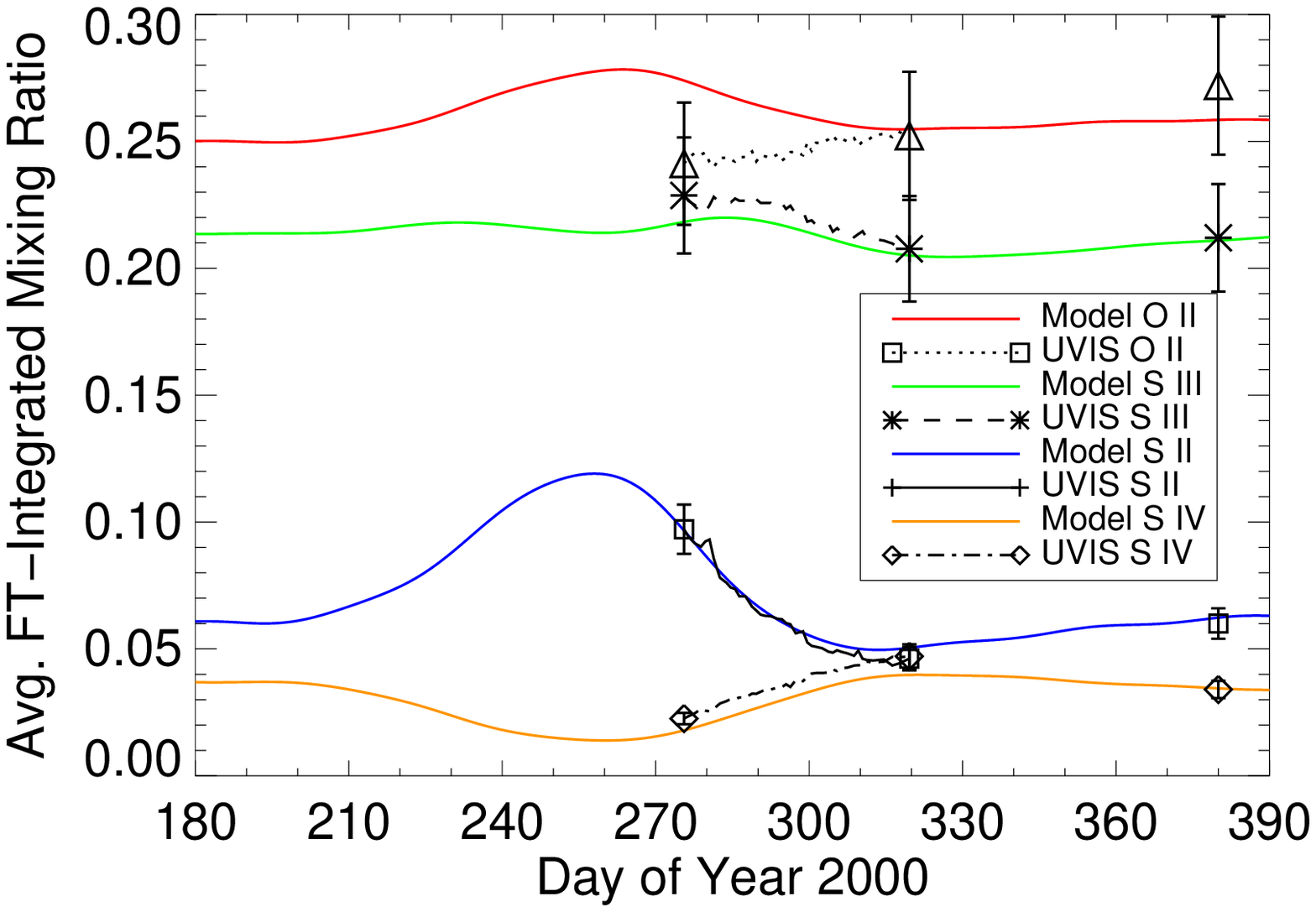}
  \caption[]{Steffl et al. \label{datavsmodel_comp} \\
    Azimuthally-averaged composition of the Io torus as observed by
    {\it Cassini} UVIS and reproduced by the time-variable azimuthal
    model. Observed mixing ratios derived from UVIS spectra are shown
    with plot symbols and the thin connecting lines. Uncertainties in
    the UVIS-derived mixing ratios are approximately 10\%, as shown by
    the error bars. Averaged mixing ratios produced by the
    time-variable azimuthal model are shown with thick solid lines.
    Although the model results shown here were produced by the
    time-variable azimuthal model, virtually identical results can be
    produced by the subcorotating hot electron model
    (Section~\ref{subcorotating_elec_model}) and the dual hot electron
    model (Section \ref{modulated_elec_model}).}
\end{figure}

\noindent {\bfseries Fig.~\ref{datavsmodel_comp} Caption} \\
Azimuthally-averaged composition of the Io torus as observed by {\it
  Cassini} UVIS and reproduced by the time-variable azimuthal model.
Observed mixing ratios derived from UVIS spectra are shown with plot
symbols and black connecting lines. Uncertainties in the UVIS-derived
mixing ratios are approximately 10\%, as shown by the error bars.
Averaged mixing ratios produced by the time-variable azimuthal model
are shown with thick solid lines.  Although the model results shown
here were produced by the time-variable azimuthal model, virtually
identical results can be produced by the subcorotating hot electron
model (Section~\ref{subcorotating_elec_model}) and the dual hot
electron model (Section \ref{modulated_elec_model}).
\\

\begin{figure}
  \includegraphics[scale=.60]{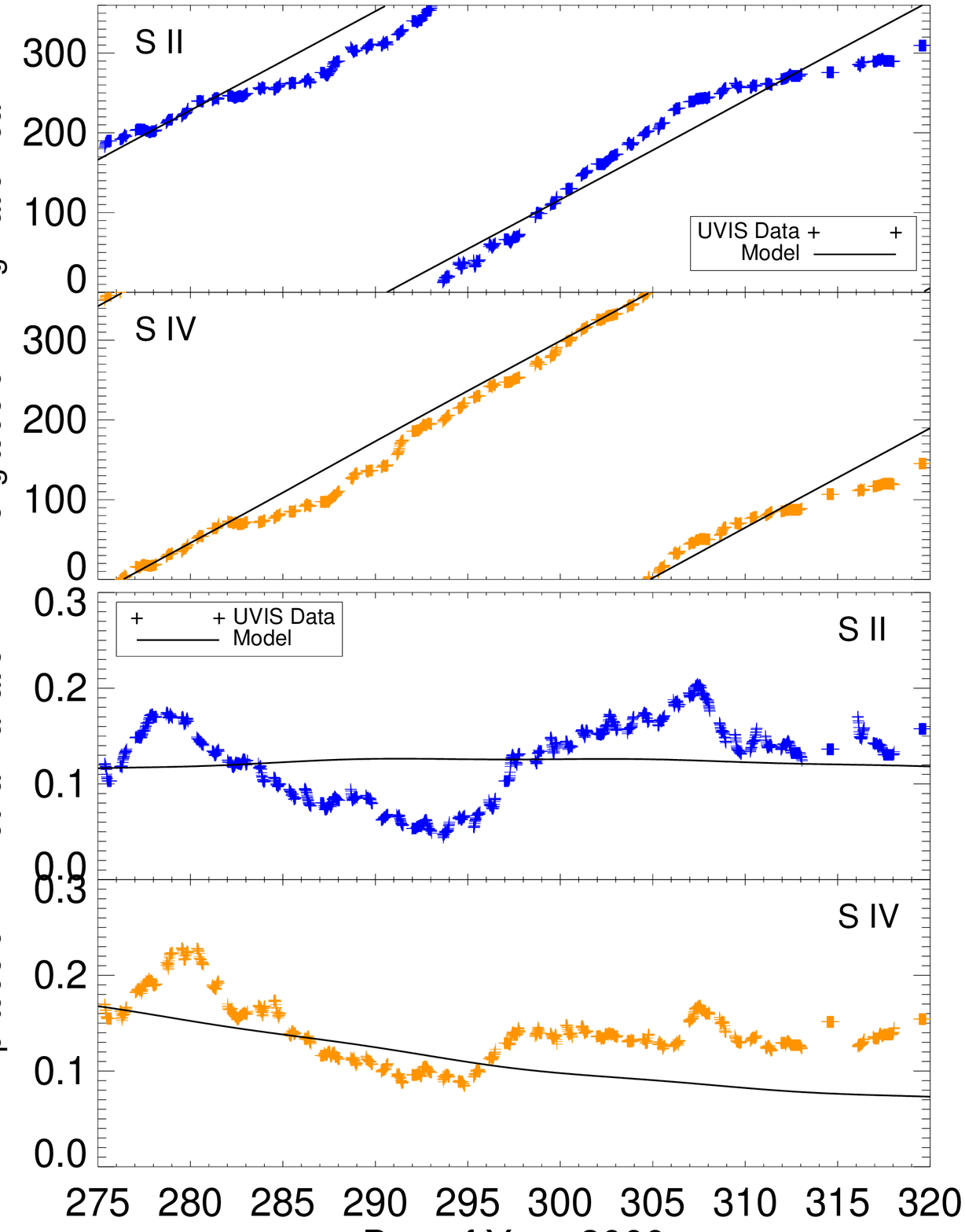}
  \caption[]{Steffl et al.  \label{datavsmodel_pa_nosys3} \\
    Comparison of {\it Cassini} UVIS data with output from the
    subcorotating hot electron model. UVIS observations of both dawn
    and dusk ansae have been averaged together. The top panel shows
    the azimuthal location (in System III coordinates) of the peak
    mixing ratios of \ion{S}{2} and \ion{S}{4}. The bottom panel shows
    the amplitude of \ion{S}{2} and \ion{S}{4} azimuthal variations.}
\end{figure}

\noindent {\bfseries Fig.~\ref{datavsmodel_pa_nosys3} Caption} \\
Comparison of {\it Cassini} UVIS data with output from the
subcorotating hot electron model. UVIS observations of both dawn and
dusk ansae have been averaged together. The top panel shows the
azimuthal location (in System III coordinates) of the peak mixing
ratios of \ion{S}{2} and \ion{S}{4}. The bottom panel shows the
amplitude of \ion{S}{2} and \ion{S}{4} azimuthal variations.
\\

\begin{figure}
  \includegraphics[scale=.60]{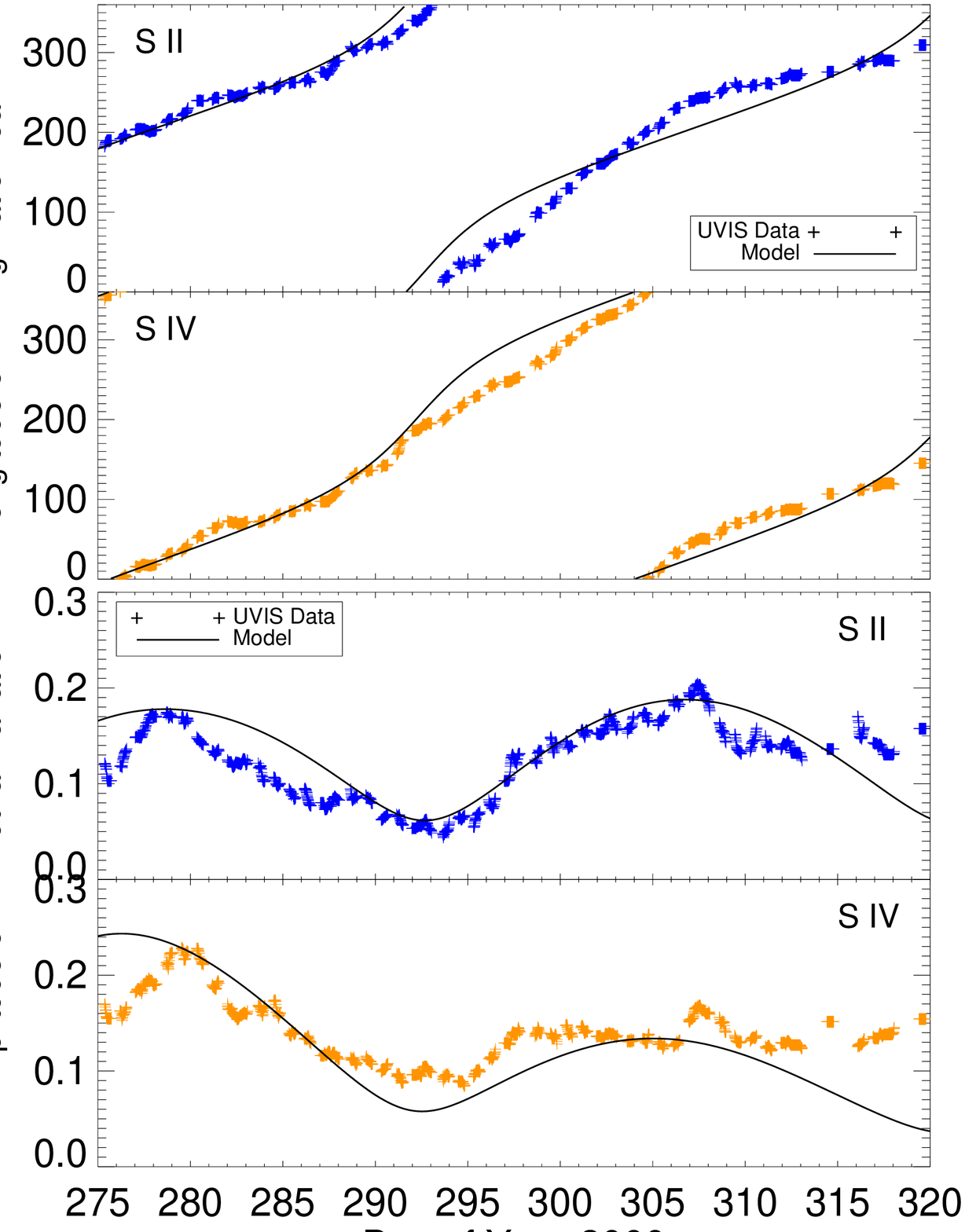}
  \caption[]{Steffl et al.  \label{datavsmodel_pa} \\ 
    Comparison of {\it Cassini} UVIS data with output from the dual
    hot electron model. This model features the superposition of two
    azimuthal variations in hot electron fraction, one with a rotation
    period of 10.07 hours and the other with the System III rotation
    period of 9.925 hours.  Unlike the azimuthal variations produced
    by the subcorotating hot electron model
    (Fig~\ref{datavsmodel_pa_nosys3}) the phase of the azimuthal
    variation increases more rapidly when the amplitude is near its
    minimum value (near DOY 293). The bottom panel shows the amplitude
    of azimuthal variations of \ion{S}{2} and \ion{S}{4}. The
    interference of the two hot electron variations creates a beat
    period of 28.8 days.}
\end{figure}

\noindent {\bfseries Fig.~\ref{datavsmodel_pa} Caption} \\
Comparison of {\it Cassini} UVIS data with output from the dual hot
electron model. This model features the superposition of two azimuthal
variations in hot electron fraction, one with a rotation period of
10.07 hours and the other with the System III rotation period of 9.925
hours.  Unlike the azimuthal variations produced by the subcorotating
hot electron model (Fig~\ref{datavsmodel_pa_nosys3}) the phase of the
azimuthal variation increases more rapidly when the amplitude is near
its minimum value (near DOY 293). The bottom panel shows the amplitude
of azimuthal variations of \ion{S}{2} and \ion{S}{4}. The interference
of the two hot electron variations creates a beat period of 28.8 days.
\\

\begin{figure}
  \includegraphics[scale=.8]{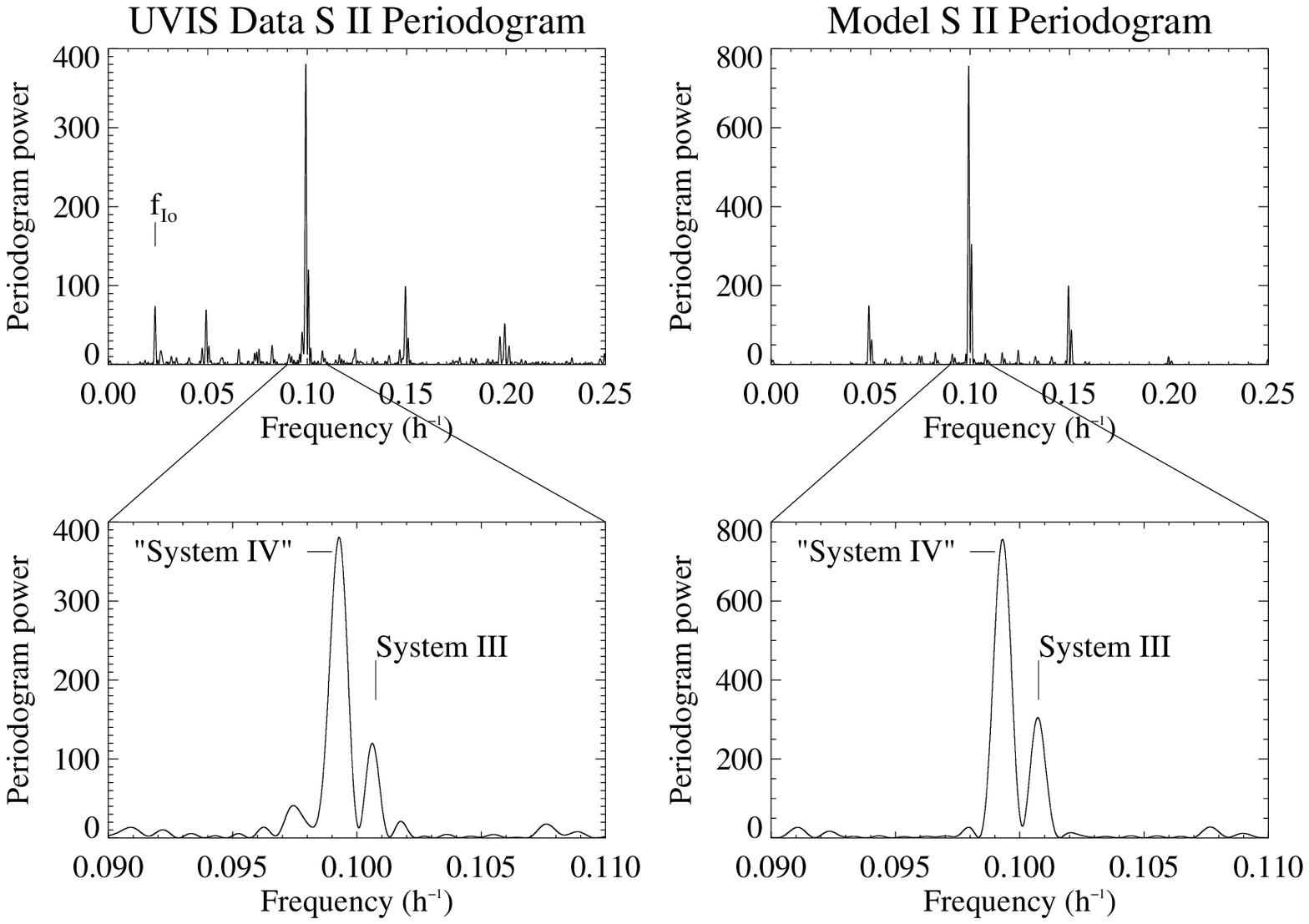}
  \caption[]{Steffl et al.  \label{datavsmodel_periodogram} \\
    Lomb-Scargle periodograms of the mixing ratio of \ion{S}{2} as
    derived from {\it Cassini} UVIS data (left) and the dual hot
    electron model (right). Model mixing ratios have been sampled at
    the same time and spatial location as the UVIS observations. Both
    data and model periodograms show a sharp peak at a frequency of
    0.0993~h$^{-1}$, corresponding to the 10.07-hour ``System IV''
    period observed by {\it Cassini} UVIS and a secondary peak at the
    System III rotational frequency. Sidebands of the two peaks can be
    seen near 0.05~h$^{-1}$ and 0.20~h$^{-1}$.  Small spurious peaks
    due to the sampling interval of the UVIS data are also present.}
\end{figure}

\noindent {\bfseries Fig.~\ref{datavsmodel_periodogram} Caption} \\
Lomb-Scargle periodograms of the mixing ratio of \ion{S}{2} as derived
from {\it Cassini} UVIS data (left) and the dual hot electron model
(right). Model mixing ratios have been sampled at the same time and
spatial location as the UVIS observations. Both data and model
periodograms show a sharp peak at a frequency of 0.0993~h$^{-1}$,
corresponding to the 10.07-hour ``System IV'' period observed by {\it
  Cassini} UVIS and a secondary peak at the System III rotational
frequency. Sidebands of the two peaks can be seen near 0.05~h$^{-1}$
and 0.20~h$^{-1}$. Small spurious peaks due to the sampling interval
of the UVIS data are also present.
\\

\begin{figure}
  \includegraphics[scale=.5]{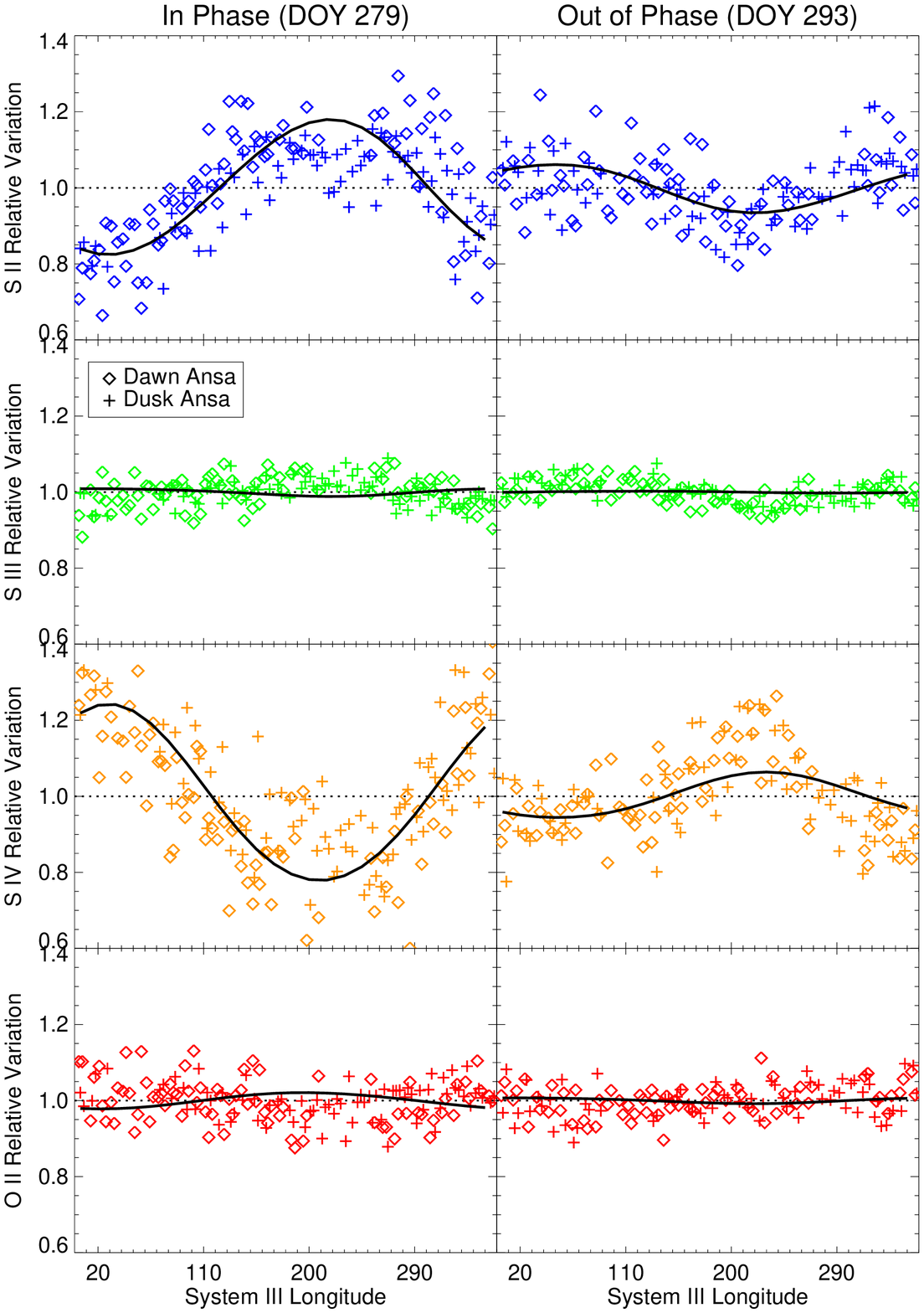}
  \caption[]{Steffl et al.  \label{datavsmodel_lonprofile} \\
    Azimuthal variation of ion mixing ratios in the Io plasma torus
    during two 2-day periods. Plotting symbols represent ion mixing
    ratios derived from {\it Cassini} UVIS data: diamonds from the
    dawn ansa and pluses from the dusk ansa. The solid lines are
    output from the dual hot electron variation model.}
\end{figure}

\noindent {\bfseries Fig.~\ref{datavsmodel_lonprofile} Caption} \\
Azimuthal variation of ion mixing ratios in the Io plasma torus during
two 2-day periods. Plotting symbols represent ion mixing ratios
derived from {\it Cassini} UVIS data: diamonds from the dawn ansa and
pluses from the dusk ansa. The solid lines are output from the dual
hot electron model.
\\

\clearpage

\end{document}